\documentclass[12pt]{article}
\usepackage{amssymb}
\usepackage{epsf}
\usepackage{epsfig}
\newcommand{\lsim}{
\mathrel{\hbox{\rlap{\hbox{\lower4pt\hbox{$\sim$}}}\hbox{$<$}}}}

\newcommand{\gsim}{
\mathrel{\hbox{\rlap{\hbox{\lower4pt\hbox{$\sim$}}}\hbox{$>$}}}}

\newcommand{\be}{\begin{equation}}
\newcommand{\ee}{\end{equation}}
\newcommand{\bi}{\begin{itemize}}
\newcommand{\ei}{\end{itemize}}
\newcommand{\ord}{{\cal O}}

\newcommand{\RE}{{\rm Re}}
\newcommand{\IM}{{\rm Im}}

\def\kpn{K^+\rightarrow\pi^+\nu\bar\nu}

\begin{document}
\begin{titlepage}

\begin{flushright}
CERN-TH/2003-200\\
TUM-HEP-526/03\\
MPP-2003-66\\
hep-ph/0309012
\end{flushright}

\vspace{0.5cm}
\begin{center}
\boldmath
\Large\bf The $B\to\pi K$ Puzzle and its Relation to\\ 
\vspace*{0.2truecm}
Rare $B$ and $K$ Decays

\unboldmath
\end{center}

\vspace{0.5cm}

\begin{center}
{\bf Andrzej J. Buras,${}^a$ Robert Fleischer,${}^b$ 
Stefan Recksiegel${}^a$ and Felix Schwab${}^{a,c}$}

\vspace{0.7truecm}

${}^a$ {\sl Physik Department, Technische Universit\"at M\"unchen,
D-85748 Garching, Germany}

\vspace{0.2truecm}

${}^b$ {\sl Theory Division, CERN, CH-1211 Geneva 23, Switzerland}

\vspace{0.2truecm}

 ${}^c$ {\sl Max-Planck-Institut f{\"u}r Physik -- Werner-Heisenberg-Institut,
 D-80805 M{\"u}nchen, Germany}

\end{center}

\vspace{0.7cm}
\begin{abstract}
\vspace{0.2cm}\noindent
The Standard-Model interpretation of the ratios of charged and neutral 
$B\to\pi K$ rates, $R_{\rm c}$ and $R_{\rm n}$, respectively, points towards
a puzzling picture. Since these observables are affected significantly by
colour-allowed electroweak (EW) penguins, this ``$B\to\pi K$ puzzle'' could 
be a manifestation of new physics in the EW penguin sector. Performing the 
analysis in the $R_{\rm n}$--$R_{\rm c}$ plane, which is very suitable for 
monitoring various effects, we demonstrate that we may, in fact, move 
straightforwardly to the experimental region in this plane through an 
enhancement of the relevant EW penguin parameter $q$. We derive analytical 
bounds for $q$ in terms of a quantity $L$, which measures the violation of 
the Lipkin sum rule, and point out that strong phases around $90^\circ$ are 
favoured by the data, in contrast to QCD factorisation. The $B\to\pi K$ 
modes imply a correlation between $q$ and the angle $\gamma$ that, in the 
limit of negligible rescattering effects and colour suppressed EW penguins, 
depends only on the value of $L$. Concentrating on a minimal flavour-violating 
new-physics scenario with enhanced $Z^0$ penguins, we find that the current 
experimental values on $B\to X_s \mu^+\mu^-$ require roughly $L\le 1.8$.
As the $B\to\pi K$ data give $L=5.7\pm2.4$, $L$ has either to move to smaller 
values once the $B\to \pi K$ data improve or new sources of flavour and CP 
violation are needed. In turn, the enhanced values of $L$ seen in the 
$B\to\pi K$ data could be accompanied by enhanced branching ratios for the
rare decays $K^+\to \pi^+\nu\bar\nu$, $K_{\rm L}\to \pi^0 e^+ e^-$, 
$B\to X_s\nu\bar\nu$ and $B_{s,d}\to \mu^+\mu^-$.
Most interesting turns out to be the correlation between the $B\to \pi K$
modes and ${\rm BR}(K^+\to \pi^+\nu\bar\nu)$, with the latter depending 
approximately on a single ``scaling" variable 
$\bar L= L\cdot (|V_{ub}/V_{cb}|/0.086)^{2.3}$.
\end{abstract}

\vfill
\noindent


%
%
%
\end{titlepage}
%
%
%

%

%
%
%
\section{Introduction}\label{sec:intro}
The rich physics potential of $B\to\pi K$ modes is attracting a lot of 
interest in the $B$-physics community \cite{BpiK-overview}. Decays of 
this kind are caused by $\overline{b}\to\overline{d}d\overline{s},\, 
\overline{u}u\overline{s}$ quark-level processes, and receive 
contributions from penguin and tree topologies, where the latter are 
associated with the angle $\gamma$ of the unitarity triangle of the 
Cabibbo--Kobayashi--Maskawa (CKM) matrix. Since the CKM factor
$|V_{us}V_{ub}^\ast/(V_{ts}V_{tb}^\ast)|\approx0.02$ is tiny, $B\to\pi K$ 
modes are governed by QCD penguins. Moreover, we have also contributions from 
electroweak (EW) penguins. In the case of $B^0_d\to\pi^-K^+$ and 
$B^+\to\pi^+K^0$ modes, these topologies are
colour-suppressed and play hence only a minor r\^ole. On the other 
hand, EW penguins contribute also in colour-allowed form to 
$B^+\to\pi^0K^+$ and $B^0_d\to\pi^0K^0$. Consequently, they are 
expected to be sizeable in these modes, i.e.\ of the same order of 
magnitude as the tree topologies. Interference between the
tree and penguin topologies leads to sensitivity on $\gamma$.

\begin{table}[h]
\vspace*{0.5truecm}
\begin{center}
\begin{tabular}{|c|c|c|c|c|}
\hline
Observable & CLEO ('03) & BaBar ('03)  & Belle ('03) & Average\\
\hline
$R$ & $1.04\pm0.26$ & $0.97\pm0.11$ & $0.91\pm0.11$ & $0.95\pm0.07$\\
$R_{\rm c}$ & $1.37\pm0.40$ & $1.28\pm0.20$ & $1.16\pm0.20$ &
$1.24\pm0.13$\\
$R_{\rm n}$ & $0.70\pm0.24$ & $0.86\pm0.15$ & $0.73\pm0.17$ &
$0.81\pm0.10$\\
\hline
\end{tabular}
\end{center}
\vspace*{-0.4truecm}
\caption{The current experimental status of the observables
$R_{\rm (c,n)}$. }\label{tab:R}
\end{table}

The isospin flavour symmetry of strong interactions suggests to consider
the following combinations of $B\to\pi K$ decays: the ``mixed'' 
$B_d\to \pi^\mp K^\pm$, $B^\pm\to \pi^\pm K$ system 
\cite{PAPIII}--\cite{defan}, the charged $B^\pm\to \pi^0 K^\pm$, 
$B^\pm\to \pi^\pm K$ system \cite{NR}--\cite{BF-neutral2}, and the neutral 
$B_d\to \pi^\mp K^\pm$, $B_d\to \pi^0 K$ system \cite{BF-neutral1,BF-neutral2}.
The CP-conserving and CP-violating observables of each system provide
sufficient information to determine $\gamma$ and a corresponding strong
phase. For the following discussion, we use the ratios of the
CP-averaged $B\to\pi K$ branching ratios introduced in \cite{BF-neutral1}:
\begin{equation}\label{R-def}
R\equiv\left[\frac{\mbox{BR}(B^0_d\to\pi^-K^+)+
\mbox{BR}(\overline{B^0_d}\to\pi^+K^-)}{\mbox{BR}(B^+\to\pi^+K^0)+
\mbox{BR}(B^-\to\pi^-\overline{K^0})}\right]\frac{\tau_{B^+}}{\tau_{B^0_d}}
\end{equation}
\begin{equation}\label{Rc-def}
R_{\rm c}\equiv2\left[\frac{\mbox{BR}(B^+\to\pi^0K^+)+
\mbox{BR}(B^-\to\pi^0K^-)}{\mbox{BR}(B^+\to\pi^+K^0)+
\mbox{BR}(B^-\to\pi^-\overline{K^0})}\right]
\end{equation}
\begin{equation}\label{Rn-def}
R_{\rm n}\equiv\frac{1}{2}\left[\frac{\mbox{BR}(B^0_d\to\pi^-K^+)+
\mbox{BR}(\overline{B^0_d}\to\pi^+K^-)}{\mbox{BR}(B^0_d\to\pi^0K^0)+
\mbox{BR}(\overline{B^0_d}\to\pi^0\overline{K^0})}\right].
\end{equation}
In Table~\ref{tab:R}, we summarise the current experimental status of these 
observables. The final averages for $R_{\rm (c,n)}$ given 
in this table have been obtained by using the average branching ratios 
from the data of CLEO \cite{CLEO03}, BaBar \cite{BaBar03} and 
Belle \cite{Belle03} that read
\be\label{branching1} 
\mbox{BR}(B^+\to\pi^0K^+)=(12.82\pm1.08)\cdot 10^{-6},
\quad
\mbox{BR}(B^+\to\pi^+K^0)=(20.62\pm1.35)\cdot 10^{-6}
\ee
and 
\be\label{branching2} 
\mbox{BR}(B^0_d\to\pi^0K^0)=(11.21\pm1.36)\cdot 10^{-6},
\quad
\mbox{BR}(B^0_d\to\pi^-K^+)=(18.16\pm0.79)\cdot 10^{-6}~.
\ee
Finally, we have used 
$\tau_{B^+}/\tau_{B_d^0}=1.086\pm 0.017$.

\begin{figure}
\vspace*{0.3truecm}
\begin{center}
\includegraphics[width=9cm]{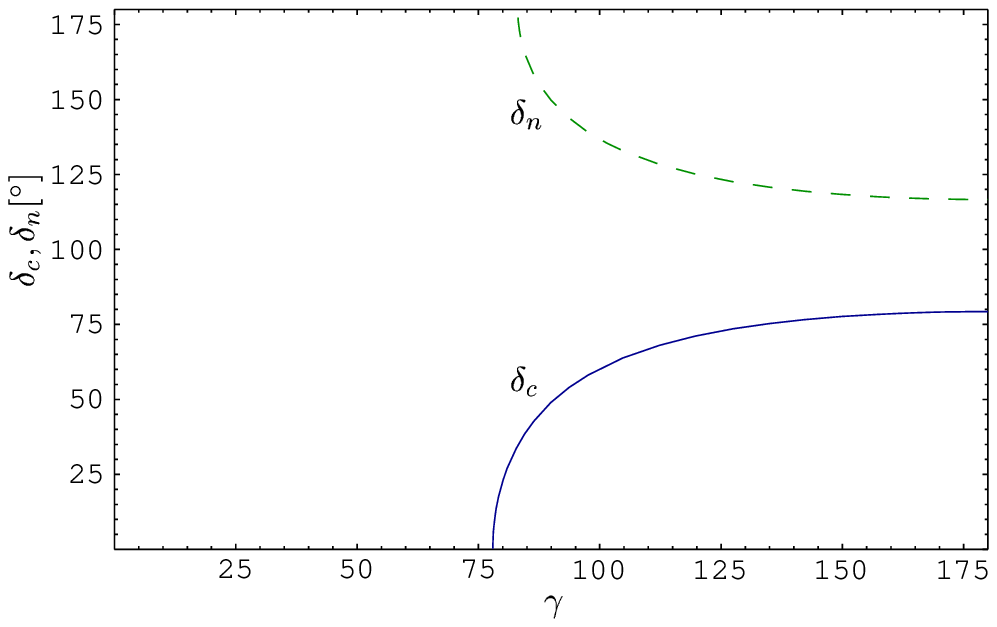}
\end{center}
\caption{$|\delta_{\rm c}|$ and $|\delta_{\rm n}|$
as functions of $\gamma$ for $R_{\rm c}=1.24$ and $R_{\rm n}=0.81$, 
respectively.}\label{fig:phases1}
\end{figure}

As was already emphasized by two of us in \cite{BF-neutral2},
the pattern of $R_{\rm c}>1$ and $R_{\rm n}<1$ is actually very
puzzling within the Standard Model (SM). To understand the problem
let us first note that $R_{\rm c}$ 
and $R_{\rm n}$ allow us to determine CP-conserving strong
phases $\delta_{\rm c}$ and $\delta_{\rm n}$ as functions of $\gamma$,
respectively. As can be seen in Fig.~\ref{fig:phases1}, the current
central values for $R_{\rm c}$ and $R_{\rm n}$ imply lower bounds for
$\gamma$ around $80^\circ$, and very different values for the strong phases. 
However, these strong phases are not expected to differ so largely from 
each other, as can be seen from their exact definitions in
\cite{BF-neutral1}. This problem becomes also obvious in the contour plots 
in the $B\to\pi K$ observable space shown in \cite{Fl-Ma}. On the other hand, 
no anomalous behaviour is indicated by the observable $R$ of the mixed 
$B\to\pi K$ system, where EW penguins only contribute in 
colour-suppressed form. Consequently, as noted in \cite{BF-neutral2}, 
this puzzle could be a manifestation of new-physics contributions in 
the EW penguin sector, which is a rather popular scenario for physics 
beyond the SM to enter the $B\to\pi K$ system \cite{FM-NP,trojan}. 
This point was also very recently re-emphasized in 
\cite{Yoshikawa:2003hb}--\cite{Beneke:2003zv}.

In 2000, when \cite{BF-neutral2} was written, the $B^0_d\to \pi^0K^0$ channel 
had just been observed by the CLEO collaboration. Now we have a much better 
experimental picture, where interestingly {\it all} three experiments point 
towards $R_{\rm c}>1$ and $R_{\rm n}<1$, as can be seen in Table~\ref{tab:R},
whereas $R\sim1$. Although the experimental uncertainties are still 
sizeable, we think that it is legitimate and interesting to return to this 
puzzle and to 
explore in more detail whether enhanced EW penguins could really provide a 
solution. Another important element of our analysis are rare $B$ and
$K$ decays. If we restrict ourselves to new-physics scenarios with
``minimal flavour violation'' (MFV) \cite{MFV} and enhanced $Z^0$ penguins, 
we obtain 
a nice connection between the $B\to\pi K$ puzzle and $B\to X_s \mu^+\mu^-$, 
$K^+\to \pi^+\nu\bar\nu$, $K_{\rm L}\to \pi^0 e^+ e^-$, 
$B\to X_s\nu\bar\nu$ and $B_{s,d}\to \mu^+\mu^-$ 
 decays. In order
to make our findings more transparent, we shall neglect colour-suppressed 
EW penguins, $SU(3)$-breaking contributions and rescattering effects. 
A more detailed analysis including these effects and addressing
more technical aspects 
can be found in \cite{BFRS03}. The outline of the present paper is as 
follows: in Section~\ref{sec:BpiK}, we explore the impact of enhanced EW 
penguins on the observables $R_{\rm c}$ and $R_{\rm n}$. In 
Section~\ref{sec:q-Z}, we discuss the connection between the value 
of the relevant $B\to\pi K$ EW penguin parameter $q$ and $Z$ penguins in 
the restricted class of MFV models specified above. These results are then 
applied in Section~\ref{sec:rare} to analyse rare $B$ and $K$ decays and
to explore the implications of the corresponding experimental constraints 
for the $B\to\pi K$ system. Finally, we summarise our conclusions 
in Section~\ref{sec:concl}.

\boldmath
\section{Enhanced EW Penguins in $B\to \pi K$ Decays}\label{sec:BpiK}
\unboldmath
If we employ the parametrisation introduced in \cite{BF-neutral1},
we may write within the approximations stated above
\begin{equation}\label{R-central}
R_{\rm c,n}=1+2 r_{\rm c,n} B \cos\delta_{\rm c,n} +
[B^2+\sin^2\gamma]r_{\rm c,n}^2,
\end{equation}
where
\begin{equation}\label{B-def}
B\equiv q-\cos\gamma
\end{equation}
is a ``universal'' quantity for the charged and the neutral $B\to\pi K$ 
systems. In addition to $\gamma$, it depends on a parameter $q$, which 
measures the ratio of the sum of the colour-allowed and colour-suppressed 
EW penguins with respect to the sum $T+C$ of the colour-allowed and 
colour-suppressed tree-diagram-like contributions. Using 
$SU(3)$ flavour-symmetry arguments, we can calculate the EW penguin 
parameter $q$ within the SM as follows \cite{NR}:
\begin{equation}\label{q-def}
\left.q\right|_{\rm SM}=0.69\times\left[\frac{0.086}{|V_{ub}/V_{cb}|}\right],
\end{equation}
where $|V_{ub}/V_{cb}|=0.086\pm 0.008$.
Here we have taken NLO corrections into account and used the most recent 
input parameters \cite{Battaglia:2003in}.
The strong phase $\omega$ associated with $q$ vanishes in the $SU(3)$ limit, 
and we have already used $\omega=0$ in (\ref{R-central}). Even values
of $\omega$ up to $20^\circ$ have very little influence on our analysis
(see \cite{BFRS03} for a detailed discussion).
  
The parameters $r_{\rm c}$ and $r_{\rm n}$ describe, 
roughly speaking, the ratio of $T+C$ and penguin amplitudes, where 
the latter are determined by the CP-averaged $B^\pm\to \pi^\pm K$ and 
$B_d\to \pi K$ rates, respectively. Using the exact definitions given in 
\cite{BF-neutral1} and taking into account that $|T+C|$ can be fixed
through the $SU(3)$ flavour symmetry with the help of the CP-averaged
$B^\pm\to\pi^\pm\pi^0$ rate \cite{Gronau:1994bn}, we arrive at 
\be\label{rc-def}
r_{\rm c}=\sqrt{2}\left|\frac{V_{us}}{V_{ud}}\right|\frac{f_K}{f_\pi}
\sqrt{\frac{{\rm BR}(B^\pm\to \pi^\pm\pi^0)}{{\rm BR}(B^\pm\to \pi^\pm K^0)}}
=0.201\pm 0.017
\ee
\be\label{rn-def}
r_{\rm n}=\left|\frac{V_{us}}{V_{ud}}\right|\frac{f_K}{f_\pi}
\sqrt{\frac{{\rm BR}(B^\pm\to \pi^\pm\pi^0)}{{\rm BR}(B^0_d\to \pi^0 K^0)}}
\sqrt{\frac{\tau_{B^0_d}}{\tau_{B^+}}}
=0.185\pm 0.018,
\ee
where we have used 
${\rm BR}(B^\pm\to \pi^\pm\pi^0)=(5.3\pm 0.8)\cdot 10^{-6}$
and have taken factorisable $SU(3)$-breaking corrections into account 
through the factor $f_K / f_\pi$.

Finally, $\delta_{\rm c}$ and $\delta_{\rm n}$ measure the strong phase 
differences between the
tree amplitude $T+C$ and the $B^+\to \pi^+ K^0$ and $B_d^0\to \pi^0 K^0$ 
penguin amplitudes, respectively. As seen in (\ref{R-central}), 
with $r_{\rm c}\approx r_{\rm n}\approx 0.2$ and $\gamma$ and $q$ 
being universal quantities, there is {\it no way} to reproduce 
$R_{\rm c}=1.24\pm0.13$ and $R_{\rm n}=0.81\pm0.10$ for the 
same values of $\delta_{\rm c}$ and $\delta_{\rm n}$. This is in 
particular clear for the special case of 
$\delta_{\rm c}\approx\delta_{\rm n}\approx0$ corresponding to QCD
factorisation \cite{Beneke:2003zv,QCDF},
where one finds generally
$R_{\rm c}\simeq R_{\rm n}>1.0$. Even the inclusion of enhanced 
``charming penguins'' \cite{Ciuchini:1997hb} does not help, which 
points to a different solution to be discussed below.

\begin{figure}
\vspace*{0.3truecm}
\begin{center}
\includegraphics[width=9cm]{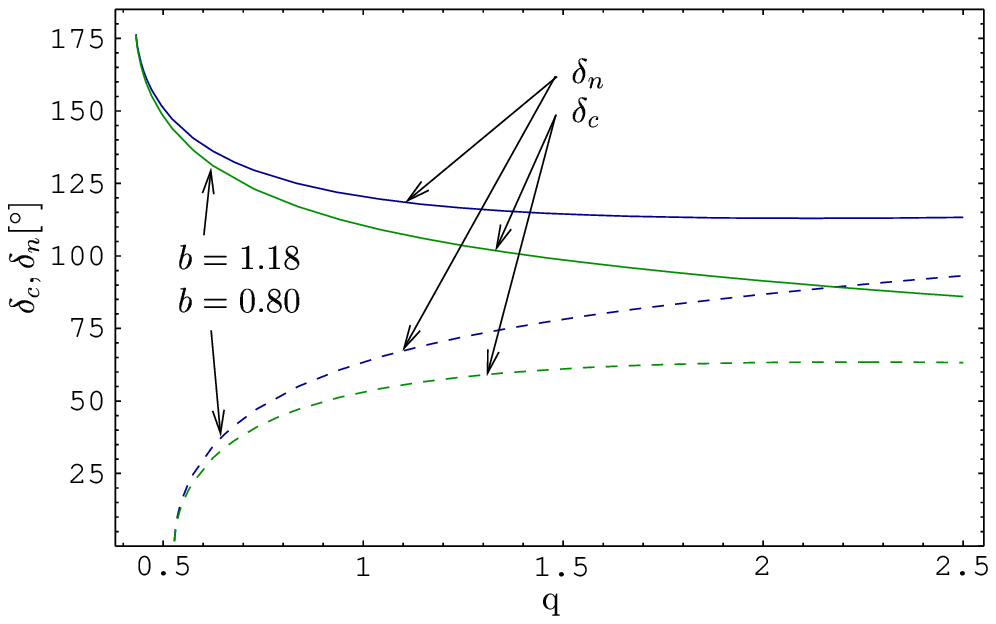}
\end{center}
\caption{$|\delta_{\rm c}|$ and $|\delta_{\rm n}|$
as functions of $q$ for $r_{\rm c}=0.20$ and $b=1.18$ (solid) and  
$b=0.80$ (dashed). 
}\label{fig:phases2}
\end{figure}

In the spirit of Fig.~\ref{fig:phases1}, the charged and neutral
$B\to\pi K$ systems were considered separately in \cite{BF-neutral2},
also in view of enhanced EW penguins. The new element we are using
here is the relation 
\begin{equation}\label{phi}
\delta_{\rm n}=\delta_{\rm c}+\varphi,
\end{equation}
where
\begin{equation}\label{phi1}
\sin\varphi=\frac{q r_{\rm c} \sin\delta_{\rm c}}{\sqrt{b}}, \qquad
\cos\varphi=\left[\frac{1-q r_{\rm c} \cos\delta_{\rm c}}{\sqrt{b}}
\right],
\end{equation}
with
\begin{equation}\label{b-def}
b\equiv \frac{R}{R_{\rm n}}=\left(\frac{r_{\rm c}}{r_{\rm n}}\right)^2
=2\left[\frac{{\mbox{BR}(B^0_d\to\pi^0K^0)+
\mbox{BR}(\overline{B^0_d}\to\pi^0\overline{K^0})}}{\mbox{BR}(B^+\to\pi^+K^0)+
\mbox{BR}(B^-\to\pi^-\overline{K^0})}\right]
\frac{\tau_{B^+}}{\tau_{B^0_d}}
=1.18\pm0.16,
\end{equation}
providing a link between the charged and neutral $B\to\pi K$ systems. 
As discussed in detail in 
\cite{BFRS03}, these relations can be derived with the help of the general 
parameterisations introduced in \cite{BF-neutral1}. The remarkable feature 
of (\ref{phi}) and (\ref{phi1}) is that we may induce a difference 
between $\delta_{\rm c}$ and $\delta_{\rm n}$ through the EW penguin 
parameter $q$, provided $\delta_{\rm c}$ is sizeable. Using the expression 
on the left-hand side of (\ref{phi1}), we obtain
\begin{equation}
\left.\sin\varphi\right|_{\rm SM}\lsim qr_{\rm c} \approx 0.14, 
\end{equation}
corresponding to a phase difference $\varphi$ of at most $\sim8^\circ$ for
$\delta_{\rm c}=90^\circ$ within the SM. This feature is the origin of 
the puzzle reflected by Fig.~\ref{fig:phases1}. However, we observe also
that the phase shift is increased through an enhancement of the
EW penguin parameter $q$. The burning question is now whether this
mechanism can actually reproduce the experimental pattern of the
observables $R_{\rm c,n}$. Before we address this exciting issue, let
us first note that the relations given in (\ref{phi1}) imply, furthermore,
the following expression:
\be\label{deltac}
\cos\delta_{\rm c}=\frac{1-b+q^2 r_{\rm c}^2}{2 q r_{\rm c}},
\ee
allowing us to calculate $\delta_{\rm c}$ as a function of $q$
for given values of $b$ and $r_{\rm c}$, which are fixed through experiment. 
In Fig.~\ref{fig:phases2}, 
the solid lines give  $\delta_{\rm c}$ and $\delta_{\rm n}$ as functions 
of $q$ for central values of $b$ and $r_{\rm c}$.

The variable $b$ coincides with $R_{00}$ in \cite{Beneke:2003zv}, and 
consequently $b=0.79\pm0.08$ in the QCD factorisation 
approach \cite{Beneke:2003zv,QCDF}, which is significantly below the 
experimental value in (\ref{b-def}). In Fig.~\ref{fig:phases2}, 
the dashed lines give  $\delta_{\rm c}$ and $\delta_{\rm n}$ as functions 
of $q$ for this low value of $b$. The 
smallness of $b$ in the latter approach can be attributed  to the 
{\it destructive} interference of QCD and EW penguin contributions 
in the $B^0_d\to\pi^0K^0 $ decay that takes place when the relevant 
phase $\delta_{\rm n}$ is small. Our finding in \cite{BF-neutral2} 
and in  Fig.~\ref{fig:phases2} that $\delta_{\rm n}> 90^\circ$ 
can be interpreted  as a {\it constructive} interference  
of QCD and EW penguin contributions 
in the $B^0_d\to\pi^0K^0 $ decay making the corresponding rate 
significantly larger than in the QCD factorisation approach.

If we now go back to (\ref{R-central}), it is an easy exercise to
derive the following expressions:
\begin{equation}\label{Rc-EXPR}
R_{\rm c}=1+\frac{B}{q}(1-b)+\left[B(B+q)+\sin^2\gamma\right]r_{\rm c}^2
\end{equation}
\begin{equation}\label{Rn-EXPR}
R_{\rm n}=1+\frac{B}{bq}(1-b)+\left[B(B-q)+\sin^2\gamma\right]
\frac{r_{\rm c}^2}{b}.
\end{equation}
Let us emphasize that (\ref{Rc-EXPR}) and (\ref{Rn-EXPR}) do {\it not} 
involve any CP-conserving strong phases. Since $b$ and $r_{\rm c,n}$ 
(up to non-factorisable $SU(3)$-breaking corrections)
can be directly determined from experiment, our two key observables 
$R_{\rm c}$ and $R_{\rm n}$ depend now only on the two ``unknowns''
$q$ and $\gamma$. Consequently, we have sufficient information to 
determine these quantities, which through (\ref{phi}),  (\ref{phi1})
and (\ref{deltac}) 
would then fix the strong phases
$\delta_{\rm c,n}$ as well. It is easy to see that (\ref{Rc-EXPR}) and 
(\ref{Rn-EXPR}) are invariant under the following transformations:
\begin{equation}
q\to -q \quad\mbox{and}\quad \gamma\to \pi-\gamma.
\end{equation}
Consequently, for each solution $(q,\gamma)$ of our problem there is
a second one, which is given by $(-q,\pi-\gamma)$. Since $q$ may, in 
principle, also be negative in the presence of new physics,
we cannot discard this case. 
However, as we will see in Section~\ref{sec:rare}, at least for
MFV models, $q>0$ turns out to be more interesting.

\begin{figure}
\begin{center}
\includegraphics[width=15cm]{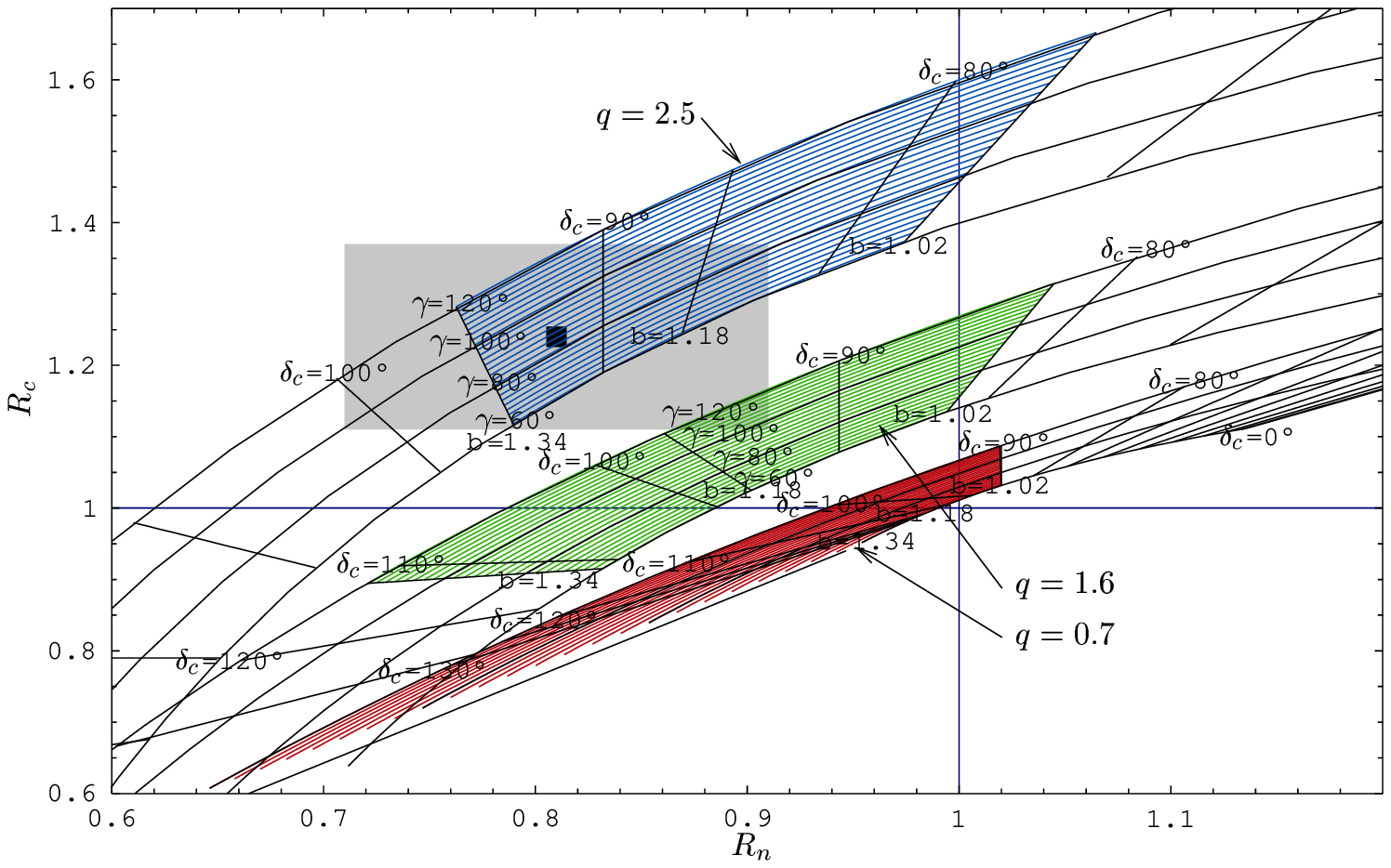}
\end{center}
\caption{Allowed regions in the $R_{\rm n}$--$R_{\rm c}$ plane for 
$q=0.7$ (SM), 1.6 and 2.5. The grey rectangle
indicates the 1-$\sigma$ experimental bounds on $R_{\rm n}$ and $R_{\rm c}$
with their central values.}\label{RcRnPlot}
\end{figure}

It is very instructive to consider the 
situation in the $R_{\rm n}$--$R_{\rm c}$ plane, as shown in 
Fig.~\ref{RcRnPlot}, where the 1-$\sigma$ experimental ranges for $R_{\rm c}$ 
and $R_{\rm n}$ are indicated by the grey rectangle. 
Each of the broad bands in 
this plane represents a given value of $q$, while different lines within 
a band correspond to different values of the angle $\gamma$, which we
vary between $60^\circ$ and $120^\circ$. A specific position on a line
fixed by $q$ and $\gamma$ corresponds to a particular value of $b$ (as 
indicated on the inside of the bands) and at the same time (through 
(\ref{deltac})) to a particular value of $\delta_{\rm c}$ (as indicated on 
the outside of the bands). The closely spaced lines are only drawn
when $b$ lies within the 1-$\sigma$ experimental region (\ref{b-def}), 
for $b$ outside this region only the ``skeleton'' of the band is drawn.

We observe two remarkable features already advertised above:
\begin{itemize}
\item 
An increase 
of $q$ brings us straightforwardly to the experimental region. This
is, in fact, necessary for {\em any} value of $\gamma$ if current data
are confirmed when precision improves.

\item 
A large strong phase $\delta_{\rm c}$ around $90^\circ$
and consequently also large $\delta_{\rm n}$ are required. 
\end{itemize}

The last feature indicates that the corrections to factorisation are 
significantly larger than estimated in the QCD factorisation approach 
\cite{Beneke:2003zv,QCDF}.
 Interestingly, evidence for a large strong
phase $\delta_{\rm c}\sim90^\circ$ follows also from an analysis of 
CP violation in $B_d\to\pi^+\pi^-$ \cite{Fl-Ma}, where the favoured
experimental sign of the corresponding direct CP asymmetry points,
for $\gamma\in [0^\circ,180^\circ]$, towards the interval 
$\delta_{\rm c}\in [0^\circ,180^\circ]$. In the decays employed in 
\cite{Fl-Ma}, EW penguins may only contribute in colour-suppressed
form.

In order to obtain further insights, it is useful to exploit that
(\ref{Rc-EXPR}) and (\ref{Rn-EXPR}) imply
\begin{equation}\label{a-central}
L\equiv \frac{(R_{\rm c}-1)+b(1-R_{\rm n})}{2 r_{\rm c}^2} = Bq,
\end{equation}
where $L$ can be determined from experiment (up to non-factorisable
$SU(3)$-breaking corrections entering through the parameter $r_{\rm c}$).
Taking into account (\ref{B-def}), this quantity allows us to calculate 
$q$ as a function of $\gamma$ with the help of
\begin{equation}\label{q-gam-exact}
q=\frac{1}{2}\left[\cos\gamma\pm\sqrt{\cos^2\gamma+4L}\right],
\end{equation}
where the plus and minus signs give $q>0$ and $q<0$, respectively. 
We observe then the third  remarkable feature:
\begin{itemize}
\item
Whereas (\ref{Rc-EXPR}) and (\ref{Rn-EXPR}) involve four experimental 
quantities $R_{\rm n}$, $R_{\rm c}$, $b$ and $r_{\rm c}$, the 
correlation between $q$ and $\gamma$ depends on a single quantity, 
the variable $L$. 
\end{itemize}

\begin{figure}
\vspace*{0.3truecm}
\begin{center}
\includegraphics[width=9cm]{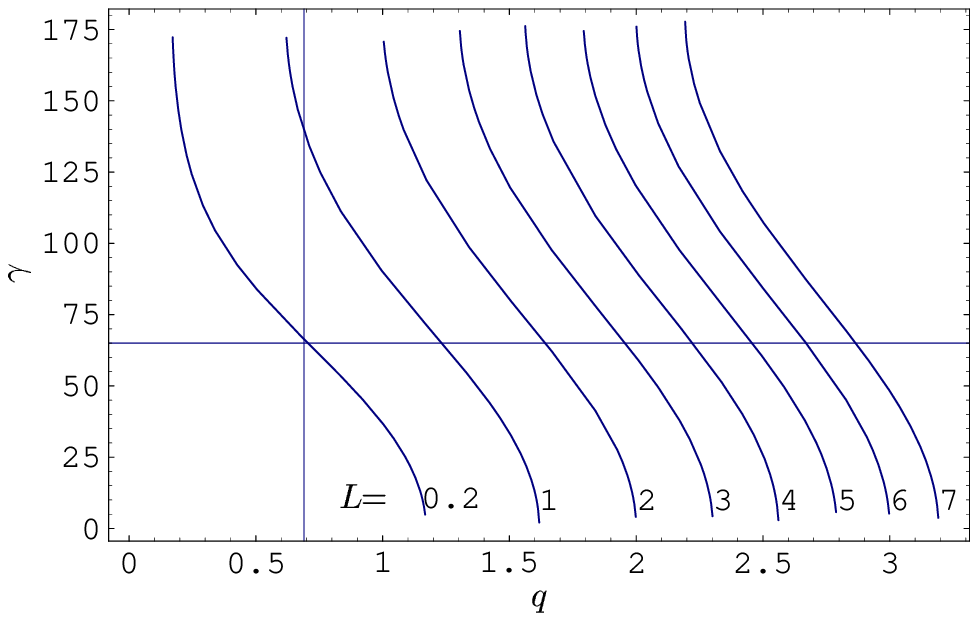}
\end{center}
\caption{$\gamma$ as a function of $q$ for different values of $L$. The 
central values of $\gamma$ and $q$ within the SM are also shown.}
\label{fig:gvsL}
\end{figure}

In Fig.~\ref{fig:gvsL}, we show
$\gamma$ as a function of $q$ for different values of $L$. The interesting 
aspect of this correlation with respect to the rare decays sensitive to 
the CKM element $|V_{td}|$ is the strong decrease of the angle $\gamma$ with 
increasing $q>0$. As the decrease of $\gamma$ is related to the decrease of 
$|V_{td}|$, this correlation has profound implications 
for the rare decay $K^+\to\pi^+\nu\bar\nu$, 
as  discussed in Section~\ref{sec:rare}.
For $q<0$, $\left|q\right|$ increases with increasing $\gamma$.

If we vary $\gamma$ between $0^\circ$ and 
$180^\circ$, we obtain
\begin{equation}
\frac{1}{2}\left[-1+\sqrt{1+4L}\right]\leq |q| \leq 
\frac{1}{2}\left[+1+\sqrt{1+4L}\right],
\end{equation}
providing interesting analytical bounds on $q$. Using the current experimental 
values given in Table~\ref{tab:R}, (\ref{rc-def}) and (\ref{b-def}), 
 we find
\be\label{qbound}
L=5.7\pm 2.4, \qquad 1.4\le |q| \le 3.4.
\ee
Consequently, the $B\to\pi K$ data favour values of $|q|$ which are 
substantially larger than the SM value in (\ref{q-def}). 
The variable $L$ measures, up to an overall factor, the violation of the 
Lipkin sum rule \cite{Lipkin}. Indeed, using the definition of
$L$ in (\ref{a-central}), we find
\be
L=\frac{2[\Gamma(B^\pm\to\pi^0K^\pm)+\Gamma(B_d\to\pi^0K)]-
[\Gamma(B^\pm\to\pi^\pm K)+\Gamma(B_d\to\pi^\mp K^\pm)]}{2 r^2_{\rm c}\Gamma
(B^\pm\to\pi^\pm K)}.
\ee
As can be seen in  (\ref{a-central}), the large value for $L$ implied by the 
$B\to \pi K$ data is directly related to $R_{\rm c}>1$ and $R_{\rm n}<1$,
i.e.\ to the $B\to\pi K$ puzzle already pointed out in \cite{BF-neutral2}.
The simple expression for $L$ in (\ref{a-central}) implies also
\begin{equation}
\left. L\right|_{\rm SM}\sim0.18,
\end{equation}
where we have used the SM values of $q=0.69$ and $\gamma=65^\circ$. The 
possible large violation of the Lipkin sum rule and theoretical 
interpretations were also discussed in \cite{Gronau:2003kj,matias} and 
very recently in \cite{Beneke:2003zv}. 
In Section~\ref{sec:rare}, we will see that $L$ provides an interesting 
link between the $B\to\pi K$ puzzle and rare $B$ and $K$ decays. 

As we already mentioned after formulae (\ref{Rc-EXPR}) and (\ref{Rn-EXPR}), 
the four quantities $q$, $\gamma$, $\delta_{\rm c}$ and $\delta_{\rm n}$
can be determined within the approximations used in this paper. The 
formulae (\ref{phi}), (\ref{phi1}), (\ref{deltac}), (\ref{Rc-EXPR}) and 
(\ref{Rn-EXPR}) are the basis for this determination and have been used 
in the plot in Fig.~\ref{RcRnPlot}. Still it is instructive 
to discuss the determination of $q$ and $\gamma$ 
in more detail. To this end, we use (\ref{B-def}) and (\ref{a-central}) 
to find
\begin{equation}\label{cosg}
\cos\gamma=q-\frac{L}{q},
\end{equation}
which allows us to eliminate $\gamma$ in the expression 
(\ref{Rn-EXPR}) for $R_{\rm n}$, thereby yielding 
\begin{equation}\label{q-extr}
q^2=U\pm\sqrt{U^2-V},
\end{equation}
with
\begin{equation}
U=\frac{b(1-R_{\rm n})+(1+L)r_{\rm c}^2}{2r_{\rm c}^2},\quad
V=\frac{(b-1)L}{r_{\rm c}^2}.
\end{equation}
Formulae (\ref{cosg}) and (\ref{q-extr}) then give the analytic expressions 
for $\gamma$ and $q$, respectively.

This discussion shows that the whole system is rather constrained. 
Consequently, the fact that a simple change of $q$ provides a solution 
to the $B\to\pi K$ puzzle is non-trivial. As is already evident from 
Fig.~\ref{RcRnPlot}, if the parameter $b$ would substantially differ from 
the experimental value in (\ref{b-def}), for instance if it was smaller than 
$1.0$,\footnote{For these considerations it is important that there are four 
independent observables, corresponding to four branching ratios; thus, $b$ 
is indeed independent of $R_{\rm n}$ and $R_{\rm c}$.} we would miss the 
experimental values of $R_{\rm n}$ and $R_{\rm c}$ even with increased $q$, 
although we could well accommodate the violation of the Lipkin sum rule. This 
feature is also reflected by the fact that (\ref{q-extr}) may not provide a 
solution at all. However, with the current experimental uncertainties, no 
discrepancy emerges and we consider it very non-trivial to be able to move 
to the experimental region in the $R_{\rm n}$--$R_{\rm c}$ plane by just
increasing the value of $q$. 

Another important consequence of our analysis are potentially large values 
of the pseudo-asymmetries introduced in \cite{BF-neutral1}, which may be
written -- within the approximations employed above -- as follows:
\begin{equation}
A_0^{\rm (c,n)}=2 r_{\rm c,n}\sin\delta_{\rm c,n}\sin\gamma.
\end{equation}
In the case of large CP-conserving phases $\delta_{\rm c,n}$ as indicated
by our numerical studies, these asymmetries could be as large as $0.3$. 
A similar pattern emerges if one assumes enhanced ``charming penguin'' 
contributions \cite{Ciuchini:1997hb}. In view of very large 
experimental uncertainties in $A_0^{\rm (c,n)}$, such large values 
of $A_0^{\rm (c,n)}$ cannot be ruled out at present.
We will briefly return to this point in Section~\ref{sec:rare}. Finally, 
the enhanced  EW penguins would enhance other branching ratios, like the 
ones for $B_s\to\pi^0\phi$ and $B_s\to\rho^0\phi$\cite{RF-EWP3}.

To summarise, if the current data will be confirmed with increasing 
experimental precision, there are four messages from these considerations: 
\begin{itemize}
\item
The EW penguin parameter $q$ must be substantially larger than in the
SM for any value of $\gamma$.
\item
Both $\delta_{\rm c}$ and $\delta_{\rm n}$ must be large and must differ 
significantly from each other, where the difference is correlated with the 
value of $q$ as given in (\ref{phi1}).
\item
The value of $b$ must be larger than $1$ in order for the enhanced value 
of $q$ to provide a solution to the $R_{\rm c}>1$ and $R_{\rm n}<1$ puzzle. 
\end{itemize}
Moreover:
\begin{itemize}
\item
A correlation between $q$ and $\gamma$ is implied by the $B\to \pi K$ 
data, as can be seen in (\ref{q-gam-exact}). It depends on the single 
variable $L$, which measures the violation of the Lipkin sum rule.
\end{itemize}

\boldmath
\section{The Value of $q$ and the $Z^0$ Penguin}\label{sec:q-Z}
\unboldmath
Next we want to investigate  whether the enhanced values of $q$ and $L$ given 
in (\ref{qbound}) are compatible with the measured branching ratios of rare 
decays in a specific extension of the SM, where a simple relation between 
the parameter $q$ entering the $B\to\pi K$ observables and the $Z^0$-penguin 
diagram function $C$ \cite{Buchalla:1995vs}, which governs many rare decays,
can be established.

To this end, we consider a simple extension of the SM, where the dominant 
new-physics contributions enter only the $Z^0$-penguin function $C$, which 
depends in the SM only on the ratio $m_t^2/M_W^2$, and equals $C\approx 0.80$. 
For this value of $C$, the $q$ given in (\ref{q-def}) is obtained.
This class of extensions of the SM has already been discussed in several
papers in the past \cite{Buras:1998ed}--\cite{Buchalla:2000sk},
but not in the context of $B\to\pi K$ decays. 
They can be considered as a restricted class of MFV models \cite{MFV} 
in which the 
CKM matrix is the only source of flavour and CP violation, the local 
operators are as in the SM and the restriction comes from the assumption 
that the dominant new physics effects enter through the $Z^0$-penguin 
diagrams.

The value of $q$  can be determined from the Wilson coefficients 
$C_9(\mu_b)$ and $C_{10}(\mu_b)$ ($\mu_b=\ord(m_b)$) of the 
$(V-A)\otimes(V-A)$ EW penguin operators $Q_9$ and $Q_{10}$ 
entering the effective Hamiltonian for $\Delta B=1$ non-leptonic decays 
\cite{Buchalla:1995vs}. 
Explicitly, in the $SU(3)$ flavour limit, one has \cite{NR}
\be
q e^{i\omega}=-\frac{3}{2}\frac{1}{\lambda |V_{ub}/V_{cb}|} 
\left[\frac{C_9(\mu_b)+C_{10}(\mu_b)}
{C_1^\prime(\mu_b)+C_2^\prime(\mu_b)}\right]\,,
\ee
where, following \cite{BF-neutral1}, we have replaced 
the Wilson coefficients $C_{1,2}$ of the current--current operators $Q_{1,2}$ 
present in the formulae of \cite{NR}  by
\be
C_1^\prime(\mu_b)=C_1(\mu_b)+\frac{3}{2}C_9(\mu_b), \qquad
C_2^\prime(\mu_b)=C_2(\mu_b)+\frac{3}{2}C_{10}(\mu_b),
\ee
as should be done in the case of enhanced EW penguins.
We observe that we have $\omega=0$ in this approximation.
  
The coefficients $C_9(\mu_b)$ and $C_{10}(\mu_b)$ can be calculated 
by means of NLO renormalisation group equations from the initial 
conditions for the Wilson coefficients at $\mu=\ord(m_t)$ entering 
the Hamiltonian in question. 
The function $C$, which appears in these initial conditions along with
box and other penguin contributions, depends on the 
gauge of the $W$ propagator, but this dependence enters only in the 
subleading terms in $m_t^2/M_W^2$ and is cancelled by the one of the 
box diagrams.

As we have seen  above, the current data for the $B\to\pi K$ decays favour 
an increased value of $q$ independently of $\gamma$ with
respect to the SM estimate, and this implies also a higher value of the 
$Z^0$-penguin function $C$.
Performing the full NLO renormalisation group analysis by means of the 
formulae in \cite{Buras:1993dy}, and assuming that only the function $C$ is 
affected by new physics, we find the following approximate but accurate 
expression for the dependence of $C$ on $q$:
\be\label{RG}
C(\bar q)= 2.35~ \bar q -0.82 , 
\qquad q=\bar q \left[\frac{0.086}{|V_{ub}/V_{cb}|}\right],
\ee 
where we have introduced $\bar q$ in order to separate the $C$ dependence
in $q$ from the $|V_{ub}/V_{cb}|$ dependence.
To our knowledge, this relation appears in the literature for the first time. 
On the other hand, in \cite{trojan}, the impact of rather involved 
new-physics scenarios that go beyond the MFV framework 
on the EW parameters  in the $B\to\pi K$ system 
has been investigated. However, these authors did not simultaneously 
discuss the correlation with rare $K$ and $B$ decays. This is the next topic 
we want to discuss.

\boldmath
\section{The Rare Decays}\label{sec:rare}
\unboldmath
The function $C(\bar q)$ is an important ingredient in any analysis of rare 
semi-leptonic $K$ and $B$ decays. Even if QCD corrections to $Z^0$ penguins 
in non-leptonic decays differ from those in the case of semi-leptonic 
decays, an explicit two-loop calculation \cite{NNLO} shows that the difference
in these corrections is very small and it is a very good approximation to use 
the same $Z^0$-penguin function in non-leptonic and semi-leptonic decays.
Moreover, these differences appear only at the NNLO level in non-leptonic 
decays \cite{NNLO}, which is clearly beyond the scope of this paper.

The present best upper bound on the function $C$ follows from 
the data on $B\to X_s l^+l^-$. A recent update of \cite{Buchalla:2000sk} 
given in \cite{Atwood:2003tg} implies that the maximal enhancement of $C$ 
over the SM value cannot be larger than 2--3, that is $ C\le 2.0$.
 Our own analysis of 
$B\to X_s l^+l^-$ indicates that indeed 
 $C>2.0$ (for $C>0$) and $|C|>2.4$ (for $C<0$) are very improbable
as they  give a BR$(B\to X_s l^+l^-)$ which is by more than a factor of
two larger than the average of the Belle and BaBar data \cite{BXll}. 
In view of substantial 
experimental and theoretical errors in the full branching ratio, it would 
be premature to attach a confidence level to these findings but in what
follows we will assume that
\be\label{C-bound}
|C|\le\left\{
\begin{array}{ll}
2.0 \,  & C>0 \\
2.4 \,  & C<0
\end{array}
\right.
\qquad
|\bar q|\le\left\{
\begin{array}{ll}
1.20 \,  & \bar q >0 \\
0.67 \,  & \bar q <0
\end{array}
\right.
\qquad
|q|\le\left\{
\begin{array}{ll}
1.47 \,  &  q >0 \\
0.82 \,  &  q <0
\end{array}
\right.
\ee
where we have used (\ref{RG}) to obtain the bounds on $|\bar q|$, 
and have then conservatively chosen $|V_{ub}/V_{cb}|\ge 0.070$
to obtain the bounds on $|q|$. These bounds can be refined in the future.
Finally, taking
$\gamma=(65\pm 10)^\circ$, as required by MFV models \cite{BPS}, we find
with the help of (\ref{a-central}) 
\be\label{L-bound}
L\le\left\{
\begin{array}{ll}
1.78\,  &  q >0 \\
1.14 \,  &  q <0
\end{array}
\right.~.
\ee

\begin{figure}
\vspace*{0.3truecm}
\begin{center}
\includegraphics[width=9cm]{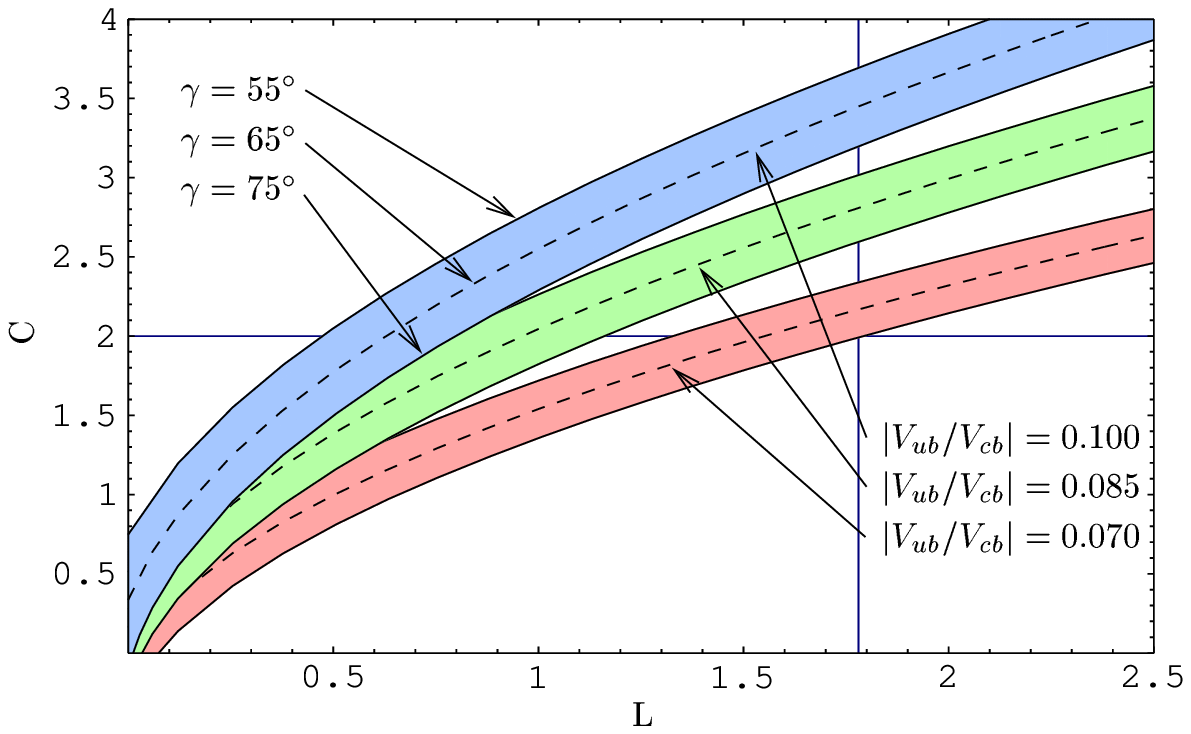}
\end{center}
\caption{$C$ as a function of $L$ for  
 $|V_{ub}/V_{cb}|=0.070,~0.085,~0.100$. 
The different lines in each band correspond to different 
$\gamma= (65\pm10)^\circ$. 
The upper bound for $L$ in (\ref{L-bound}) is obtained for 
$\gamma= 75^\circ$ and $|V_{ub}/V_{cb}|=0.070$. 
}\label{fig:CL}
\end{figure}

In Fig.~\ref{fig:CL}, we show $C$ in the case of $C>0$ as a function of
$L$ for $\gamma=(65\pm 10)^\circ$ and various values of 
$|V_{ub}/V_{cb}|$. The different lines in the $|V_{ub}/V_{cb}|$ bands
correspond to different values of $\gamma$.  
This plot establishes the connection between $B\to \pi K$ decays and rare $K$ 
and $B$ decays in the class of simple models considered here. This connection 
will become more precise when the determinations of $\gamma$ and 
$|V_{ub}/V_{cb}|$ improve.

The allowed values for $L$ in (\ref{L-bound}) 
 are clearly outside the  1-$\sigma$ range for $L$ in (\ref{qbound}) and more
than a factor of three below the central values of $L$. 
Moreover, for $\gamma= (65\pm10)^\circ$, the allowed range for $\left|q\right|$
from $B\to K\pi$ decays with $L$ given in (\ref{qbound}) is $1.6\le
 \left|q\right| \le 3.1$.
Consequently, in the context of the simple new-physics scenario considered
here, the enhancement of EW penguins implied by the $B\to\pi K$ 
data appears to be too strong to be consistent with the data on  
${\rm BR}(B\to X_s l^+l^-)$, unless values for $L$ outside the range 
(\ref{qbound}) -- but still higher than the SM value $L\approx 0.2$ --
are considered. If this really turned out to be the case, 
the enhanced 
$L$ implied by the $B\to \pi K$ data should be accompanied by an 
enhanced ${\rm BR}(B\to X_s l^+l^-)$  
close to the upper limit coming from the Belle
and BaBar data. 
Similarly, the enhanced value of $L$ should be accompanied by an enhanced 
forward--backward asymmetry in this decay that increases with increasing 
$C(\bar q)$.

In this spirit we will now analyse the $K\to \pi \nu\bar\nu$ and 
$K_L\to\pi^0 e^+ e^-$ decays 
 with the vision that, if $B\to \pi K$ decays indeed signal an 
enhancement of $L$ and consequently of $C$, this enhancement
could eventually be tested through these rare decays.

The branching ratios for $K\to \pi \nu\bar\nu$ and 
$K_L\to\pi^0 e^+ e^-$
are usually written in terms of the functions
$X$ and $Y$ \cite{Buchalla:1995vs}, respectively. These functions are 
linear combinations 
of the $C$ function and the $\Delta F =1$ box diagram function that we
assume to take the 
SM value, $B_{\rm SM}=-0.182$ for $m_t(m_t)=167~{\rm GeV}$. 
To our knowledge, this assumption is satisfied in all MFV models that have 
been considered in the literature.

We find then 
\be
X(\bar q)=2.35~\bar q -0.09, \qquad Y(\bar q)=2.35~ \bar q - 0.64,
\ee
and consequently, using (\ref{C-bound}),
\be\label{XY-bound}
|X|\le\left\{
\begin{array}{ll}
2.73 \,  & X>0 \\
1.66 \,  & X<0
\end{array}
\right.
\qquad
|Y|\le\left\{
\begin{array}{ll}
2.18 \,  & Y >0 \\
2.21 \,  & Y <0
\end{array}
\right.
\ee
to be compared with $X=1.53\pm0.04$ and $Y=0.98\pm0.04$ in the SM.

Inserting $X(\bar q)$ and $Y(\bar q)$ in the known expressions for 
the branching ratios 
of various rare decays \cite{Buchalla:1995vs}, 
it is straightforward to calculate these branching 
ratios for a given $q$ and $L$ considered in the context of 
$B\to\pi K$ decays. 
Moreover, the correlation between $q$ and the angle $\gamma$ 
in (\ref{q-gam-exact}) will also have some impact 
on the rare decays sensitive to $V_{td}$ whereas it has no impact on 
decays sensitive to $V_{ts}$.
These correlations between the new physics 
in $B\to \pi K$ and rare $K$ and $B$ decays are discussed in more detail in
\cite{BFRS03}. Below we discuss only selected aspects of this analysis.

We consider first the decays $K\to \pi \nu\bar\nu$ for which
the branching ratios are given as follows \cite{Buras:2003wd}:
\begin{equation}\label{bkpn}
{\rm BR}(\kpn)=4.75\cdot 10^{-11}\cdot\left[\left(\IM F_t \right)^2+
\left(\RE F_c +\RE F_t\right)^2 \right]~,
\end{equation}
\begin{equation}\label{bklpn}
{\rm BR}(K_{\rm L}\to\pi^0\nu\bar\nu)=2.08\cdot 10^{-10}\cdot
\left(\IM F_t  \right)^2~,
\end{equation}
where
\be 
F_c={\lambda_c\over\lambda}P_0(X), \qquad 
F_t={\lambda_t\over\lambda^5}X(\bar q)~.
\ee
Here $\lambda_i=V^\ast_{is}V_{id}$, whereas $P_0(X)=0.39\pm 0.06$ 
results from the internal charm contribution \cite{BB}, which is assumed 
not to be affected by new physics.

Note that for a given value of the angle $\gamma$, and the values of
$|V_{cb}|$ and $|V_{ub}/V_{cb}|\ge 0.070$,  
the CKM factors $\lambda_i$ can be calculated, 
and consequently we can study the branching ratios in question as a function 
of $\gamma$ and $L$. 
The $|V_{ub}/V_{cb}|$ dependence in the relation of $C$ to $L$ shown in 
Fig.~\ref{fig:CL} and the correlation 
between $C(\bar q)$ and $\gamma$ implied by (\ref{q-gam-exact}) have 
to be consistently taken into account in this analysis.

\begin{figure}
\begin{center}
\includegraphics[width=15cm]{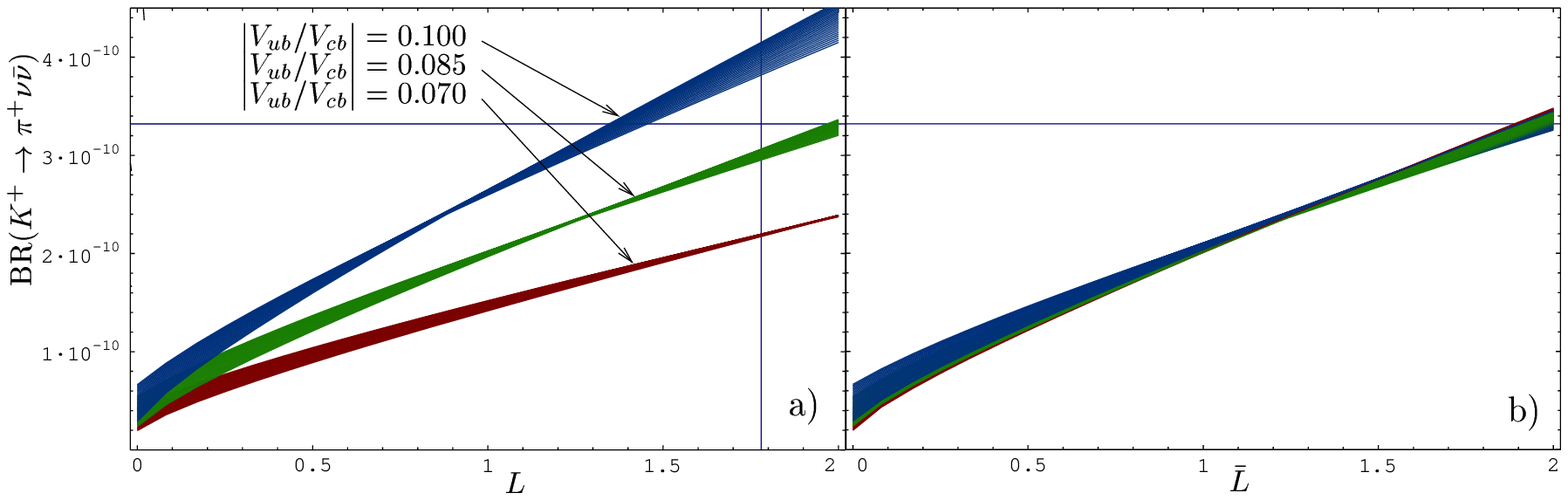}
\end{center}
\caption{ {\rm BR}($K^+\to \pi^+ \nu\bar\nu$) as a function of $L$ and 
$\bar L$  for 
$\gamma=(65\pm10)^\circ $.
 \label{Knunu}}
\end{figure}

In Fig.~\ref{Knunu}a, we show 
${\rm BR}(K^+\to \pi^+ \nu\bar\nu)$ as 
a function of $L$ for $\gamma=(65\pm10)^\circ$. 
The horizontal line represents the $68\%$ C.L. upper bound 
following from the AGS E787 collaboration result \cite{AGSE787}
\be
{\rm BR}(K^+\to\pi^+\nu\bar\nu)=(15.7^{+17.5}_{-8.2})\cdot 10^{-11}.
\ee

The sensitivity of ${\rm BR}(K^+\to\pi^+\nu\bar\nu)$ to $\gamma$ seen in
Fig.~\ref{Knunu}a is substantially smaller than in the usual SM analysis.
This feature is due to the correlation between $q$ and $\gamma$ in 
(\ref{q-gam-exact}), and the correlation between $C$ and $\gamma$ 
for fixed $L$ in Fig.~\ref{fig:CL}: the variations of $\gamma$ and $C>0$ 
in ${\rm BR}(K^+\to\pi^+\nu\bar\nu)$ compensate each other to a large 
extent.

On the other hand, we observe a strong sensitivity 
of ${\rm BR}(K^+\to\pi^+\nu\bar\nu)$ obtained in this manner 
on $|V_{ub}/V_{cb}|$, which is essentially not present in the usual 
analysis. This fact 
originates in the correlation between $C$ and $L$ in Fig.~\ref{fig:CL} 
that depends on $|V_{ub}/V_{cb}|$. However, as demonstrated in
Fig.~\ref{Knunu}b, this dependence in ${\rm BR}(K^+\to\pi^+\nu\bar\nu)$ 
can be summarised to a good approximation by introducing the following 
``scaling'' variable:
\be\label{Lbar}
\bar L \equiv L \left(\frac{|V_{ub}/V_{cb}|}{0.086}\right)^{2.3}~.
\ee
Then  ${\rm BR}(K^+\to\pi^+\nu\bar\nu)$ depends to a good approximation 
only on $\bar L$, with a weak residual dependence on 
$|V_{ub}/V_{cb}|$ and $\gamma$. Needless to say, the knowledge of 
 $|V_{ub}/V_{cb}|$ is essential for the usefulness of the 
correlation between $B\to\pi K$ and rare decays discussed here.

The bound $L\le 1.8$ required by the 
${\rm BR}(B\to X_s \mu^+\mu^-)$ data translates into $\bar L\le 1.1$ 
and consequently values for ${\rm BR}(K^+\to \pi^+ \nu\bar\nu)$
as high as $25\cdot 10^{-11}$ are allowed. 
However, even for $0.5 \le \bar L\le 1.0$,    
${\rm BR}(K^+\to \pi^+ \nu\bar\nu)$ can be a factor of 2 larger
than the SM prediction $(7.7\pm 1.2)\cdot 10^{-11}$ 
\cite{Buras:2003wd,Gino03}, 
and close to the central values of the
AGS E787 experiment. We also note that the present data on
${\rm BR}(K^+\to \pi^+ \nu\bar\nu)$ put a much weaker constraint on $L$ 
than $B\to X_s \mu^+\mu^-$, but this can change in the future as the 
$K^+\to \pi^+ \nu\bar\nu$ decay is theoretically cleaner.
Since the experimental situation 
for $K_{\rm L}\to\pi^0\nu\bar\nu$ is not as 
satisfactory, we mention only that for $\bar q=1.20$ this decay
is enhanced roughly by a factor of 3 with respect to its SM value.

Another interesting process is the rare decay $K_{\rm L}\to\pi^0 e^+e^-$ 
reconsidered recently within the SM \cite{ABI} in view 
of new NA48 data  on $K_{\rm S}\to\pi^0 e^+e^-$
and $K_{\rm L}\to \pi^0\gamma\gamma$ \cite{NA48} that allow a much better 
evaluation of the indirectly CP-violating and CP-conserving  contributions 
to ${\rm BR}(K_{\rm L}\to\pi^0 e^+e^-)$. In order to illustrate the 
implications of the enhanced $Z^0$ penguins on this ratio we set, in the 
spirit of \cite{Buras:1998ed}, all remaining loop functions at their SM values 
and keep only the function $C$ as a free parameter. Setting moreover 
all other parameters of \cite{ABI} at their central values, we find
\be\label{kpee}
{\rm BR}(K_{\rm L}\to\pi^0 e^+e^-)=10^{-12}\left[18.3+12.5 \bar y_{7V}+ 
4.4 (\bar y_{7V}^2+\bar y_{7A}^2)\right],
\ee
where
\be
\bar y_{7V}=0.56+0.69 Y(\bar q)-0.64 C(\bar q), 
\qquad \bar y_{7A}=-0.69 Y(\bar q). 
\ee
In the SM, $\bar y_{7V}=0.73$ and $\bar y_{7A}=-0.68$ at the NLO level 
\cite{BMML}.
Note that $\bar y_{7V}$ depends only very weakly on $\bar q$.

\begin{figure}
\begin{center}
\includegraphics[width=15cm]{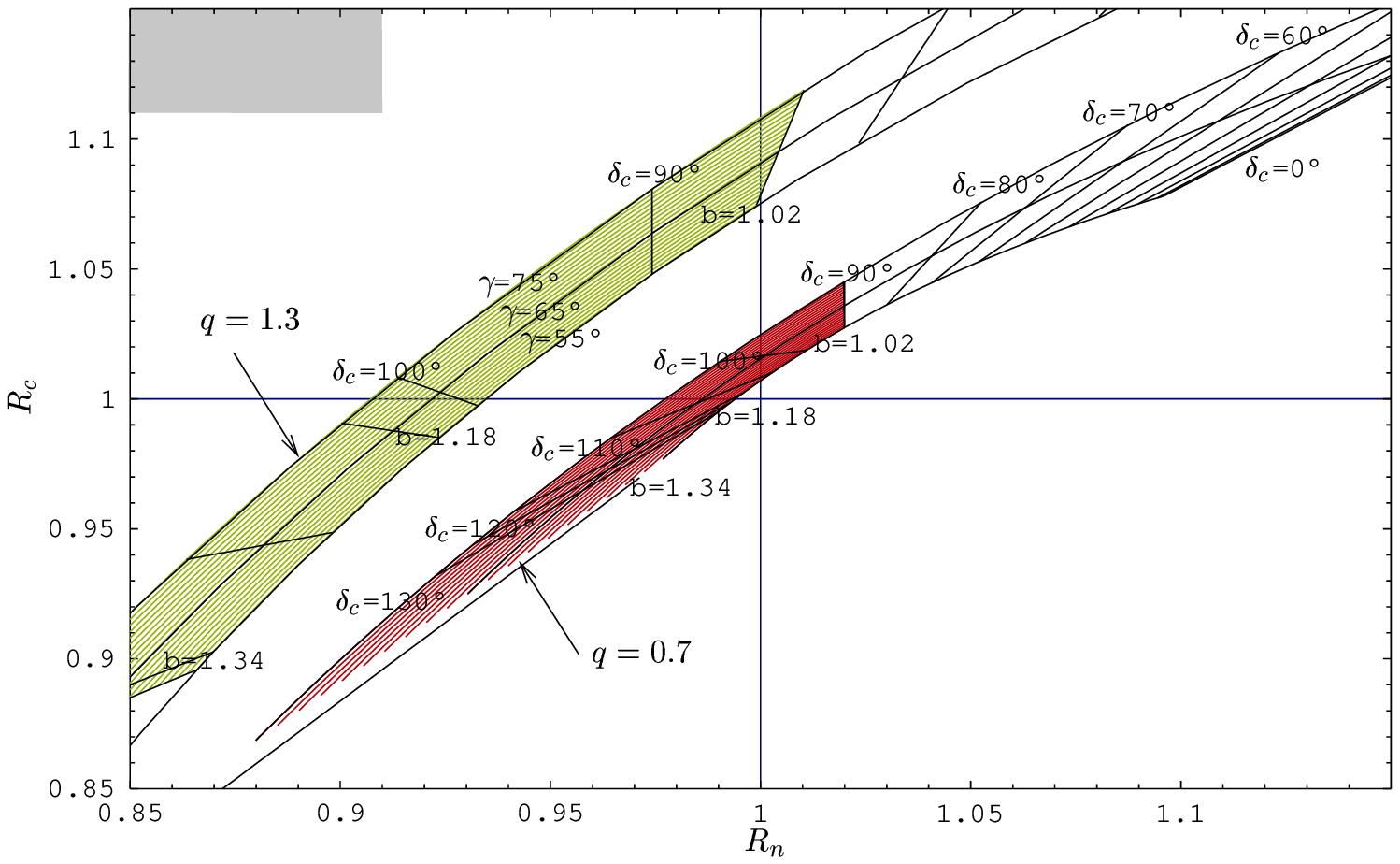}
\end{center}
\caption{Allowed regions in the $R_{\rm n}$--$R_{\rm c}$ plane for $q=0.7$ 
and $q=1.3$. The grey rectangle represents the lower right hand corner
of the experimental 1-$\sigma$ region that can be fully seen in
Fig.~\ref{RcRnPlot}. \label{RcRnM}}
\end{figure}

For ${\rm BR}(K_{\rm L}\to\pi^0 e^+e^-)$, we find then with 
$\bar q_{\rm max}=1.20$ 
the central value $4.1\cdot 10^{-11}$   
to be compared with the 
central value $3.2\cdot 10^{-11}$ within the SM \cite{ABI}
 and the experimental 
upper bound from KTeV \cite{KTEVKL}: $2.8 \cdot 10^{-10}~(90\%$C.L.). 
Even for 
$\bar q=2.5$ one finds $9.1\cdot 10^{-11}$, which is still compatible with
the data. Consequently, this decay does not offer useful bounds on
$q$ and $L$ at present.

Next, we would like to comment briefly on some other decays
that, while not yet observed at the level required to further tighten the 
bounds, do exhibit rather striking effects: these are the decays 
$B\to X_s\nu\bar\nu$ 
and $B_{s,d}\to \mu^+\mu^-$, where the values in
(\ref{XY-bound}) correspond to enhancements over the SM estimates
of the branching ratios by factors of $3.2$ and $5.0$, respectively.
This gives roughly ${\rm BR}(B\to X_s\nu\bar\nu)= 1\cdot 10^{-4}$, 
${\rm BR}(B_{s}\to \mu^+\mu^-)= 2\cdot 10^{-8}$ 
and  ${\rm BR}(B_{d}\to \mu^+\mu^-)= 5\cdot 10^{-10}$,
all compatible with the existing upper bounds on these processes.
Observation of these modes at this level could signal the presence of 
enhanced $Z^0$ penguins.

Finally, we can now ask what is the impact of the rare decays on the 
$R_{\rm n}$--$R_{\rm c}$ plane in Fig.~\ref{RcRnPlot}.
In Fig.~\ref{RcRnM}, we show the area close to $(1,1)$ in
the $R_{\rm n}$--$R_{\rm c}$ plane and plot the contours for $q=0.7$ 
and $q=1.3$ (as consistent with (\ref{C-bound})) and $\gamma=(65\pm10)^\circ$.
The comments about how variations of $q$, $b$ and $\delta_{\rm c}$ are
denoted made in the context of Fig.~\ref{RcRnPlot} still apply.

We observe that a modest shift of the experimentally allowed region
brings the values of $R_{\rm n}$, $R_{\rm c}$ and $b$ into agreement with the
scenario that employs an enhanced $C$ ($q=1.3$). A larger shift
(corresponding to shifts of each of the four $B\to K\pi$ branching ratios 
into the respective preferred direction by $1$--$1.6\,\sigma$) will even
move the experimental region towards the contour that only employs
the SM value for $q$. In both of these cases, however, the strong
phases have to be close to $90^\circ$, implying large direct CP
asymmetries and contradicting QCD factorisation. As can be seen clearly from
Fig.~\ref{RcRnM}, small values of $\delta_{\rm c}$ corresponding to small
direct CP asymmetries can only be obtained for $R_{\rm n}\gtrsim 1.1$, which 
is currently consistently disfavoured by all three experiments.
Small asymmetries are also obtained for $R_{\rm c},R_{\rm n} <1$, but in this
case $\delta_{\rm c},\delta_{\rm n} \approx 180^\circ$ 
(c.f.\ Fig.~\ref{fig:phases2}).

\section{Conclusions}\label{sec:concl}
The current data on $B\to\pi K$ decays cannot be easily explained within the
Standard Model. Extending the 2000 analysis of two of us, we have demonstrated
that the present data on these decays can be correctly described, provided:
\begin{itemize}
\item
the EW penguin parameter $q$ is by a factor of 2--4 larger than its SM value,
\item
the strong phases
$\delta_{\rm c}$ and $\delta_{\rm n}$ are large and in particular 
$\cos\delta_{\rm n}$ is negative.
\end{itemize}

These findings are true for any value of the angle $\gamma$ so that changing 
only $\gamma$ does not solve the problem if the former two conditions are 
not satisfied. Consequently
\begin{itemize}
\item
The present data indicate that the corrections to factorisation are 
significantly larger than estimated in the QCD factorisation approach and,
moreover, new-physics contributions may be signalled.
\end{itemize}

Using the general parametrisation of $B\to\pi K$ decays proposed in
\cite{BF-neutral1}, we have derived a number of new relations with 
the help of the $SU(3)$ flavour symmetry, taking factorisable $SU(3)$-breaking 
corrections into account. In particular:
\begin{itemize}
\item The $B\to\pi K$ data imply a correlation between $q$ and $\gamma$ 
which depends on a single variable $L$. This quantity measures the violation 
of the Lipkin sum rule and can be determined experimentally. In the case of
$q>0$, an increase of $\gamma$ decreases the $q$ required to fit the data. 
Moreover, $q$ increases with increasing $L$.
\item The CP-conserving strong phase difference $\delta_{\rm n}-\delta_{\rm c}$
is correlated with $q$ and $\delta_{\rm c}$, and increases with $q$ and 
$\sin\delta_{\rm c}$.
\item The measurement of the four CP-averaged $B\to\pi K$ branching ratios 
allows us to determine $q$, $\gamma$, $\delta_{\rm c}$ and $\delta_{\rm n}$.
\item Consequently, the fact that an increase of $q$ allows us to obtain 
straightforwardly an agreement with the data should be considered as
very non-trivial.
\item We have proposed to monitor these correlations in the form of allowed
regions in the $R_{\rm n}$--$R_{\rm c}$ plane. 
\end{itemize}

Concentrating then on a MFV new-physics scenario with enhanced $Z^0$ penguins,
we have derived relations between the parameter $q$ relevant for 
$B\to\pi K$ decays and the functions $C$, $X$ and $Y$ that enter 
the branching ratios for rare $K$ and $B$ decays. This allowed us to
analyse the correlations between the latter decays and the $B\to\pi K$ 
system with the following findings:
\begin{itemize}
\item In the context of the simple new-physics scenario considered
here, the enhancement of EW penguins implied by the $B\to\pi K$ 
data appears to be too strong to be consistent with the data on  
${\rm BR}(B\to X_s l^+l^-)$: whereas the latter decays imply 
$L\le 1.8$, the  $B\to\pi K$ data requires
$L=5.7\pm2.4$. Consequently,  
either $L$ has to move to smaller values once the $B\to\pi K$ 
data improve, or new sources of flavour and CP violation are needed. 
\item Enhanced values of $L$, while still smaller than those calculated
from the present $B\to\pi K$ data, could be accompanied by  
enhanced branching ratios for the rare decays $K^+\to \pi^+\nu\bar\nu$, 
$K_{\rm L}\to \pi^0 e^+ e^-$, $B\to X_s\nu\bar\nu$ and 
$B_{s,d}\to \mu^+\mu^-$.
\item In particular, we have found a correlation between the $B\to \pi K$
modes and ${\rm BR}(K^+\to \pi^+\nu\bar\nu)$, with the latter depending 
approximately only on a single ``scaling'' variable 
$\bar L= L\cdot (|V_{ub}/V_{cb}|/0.086)^{2.3}$.
For $L\approx 1.0$, an enhancement of ${\rm BR}(K^+\to \pi^+\nu\bar\nu)$ 
by a factor of two with respect to the SM is expected.
\end{itemize}

In order to describe our results in a transparent manner, we have 
neglected rescattering effects, colour-suppressed EW penguins, 
and the strong phase $\omega$, which enters the EW penguin parameter
$q$ and is predicted to vanish in the $SU(3)$ limit. These effects are
incorporated in the general parametrisation presented in \cite{BF-neutral1}
and will be discussed in detail in \cite{BFRS03}: we find that rescattering 
effects have a minor impact and cannot explain the $B\to\pi K$ puzzle.
The inclusion of colour-suppressed EW penguins makes this puzzle slightly 
more pronounced, i.e.\ the required value of $q$ increases. Larger effects 
could emerge from a non-vanishing value of $\omega$, as already discussed 
in \cite{BF-neutral1,BF-neutral2}, but as long as $\omega\le 20^\circ$, 
also these effects are small. It should be recalled that a non-vanishing 
$\omega$ could come only from non-factorisable $SU(3)$-breaking effects 
and a value substantially larger than $20^\circ$ appears to be very unlikely.

It will be very exciting in the next couple of years to follow the 
development of the values of $R_{\rm c,n}$, $R$, $r_{\rm c,n}$, $b$, $q$ and 
$\delta_{\rm c,n}$, and to monitor the allowed regions in the 
$R_{\rm n}$--$R_{\rm c}$ plane. Equally interesting will be the correlation
of the $B\to\pi K$ data with those for rare $B$ and $K$ decays for which 
more accurate measurements should soon be available.

\vspace*{0.5truecm}
{\bf Acknowledgements}
We thank Andreas Weiler for very useful discussions.
This research was partially supported by the German ``Bundesministerium f\"ur 
Bildung und Forschung'' under contract 05HT1WOA3, and by the 
``Deutsche Forschungsgemeinschaft'' (DFG) under contract Bu.706/1-2.

\end{document}